# Ultrafast non-volatile rewritable ferroaxial switching

Z. Zeng[1,2], M. Först[1], M. Fechner[1], D. Prabhakaran[2], P. G. Radaelli[2], A. Cavalleri[1,2]

[1]Max Planck Institute for the Structure and Dynamics of Matter, Hamburg, Germany
[2]Department of Physics, Clarendon Laboratory, University of Oxford, Oxford, United Kingdom

**Ultrafast switching of ferroic phases is an important research frontier, with significant technological potential. Yet, current efforts are meeting some key challenges, ranging from limited speeds in ferromagnets to intrinsic volatility of switched domains due to uncompensated depolarizing fields in ferroelectrics. Unlike these ferroic systems, ferroaxial materials host bistable states that do not break spatial-inversion or time-reversal symmetry, and are therefore immune to depolarizing fields. Yet, they are difficult to manipulate because external axial fields are not easily constructed with conventional methods. Here, we demonstrate ultrafast switching of ferroaxial order by engineering an effective axial field made up of circularly driven terahertz phonon modes. A switched ferroaxial domain remains stable for many hours and can be reversed back with a second terahertz pulse of opposite helicity. The effects demonstrated here may lead to a new platform for ultrafast information storage.**

Ferroic orders in solids exhibit pairs of energetically degenerate macroscopic states that can be switched by the application of a corresponding conjugate field. Well-known examples of ferroic order are ferromagnetism and ferroelectricity, which break time reversal and spatial inversion symmetry, respectively, and are used in memories, sensors, actuators and spintronic devices(*1, 2*). Ferroaxial order is a distinct type of ferroic order, characterized by a rotational texture that hosts vortices of electric dipoles. Ferroaxial order does not break either spatial inversion or time reversal symmetry (*3-6*). Because, no net polarization nor net magnetization are present, ferroaxials are free from depolarizing or stray fields (*7, 8*), making them promising candidates for stable, non-volatile data storage. Furthermore, ferroaxial order provides a symmetry channel that mediates the coupling between chiral magnetic textures and electric polarizations (*9-12*), and is of interest to research in multiferroics. These unique functional properties have driven rapid advances in both the exploration of new ferroaxial materials (*13-18*) and the development of new techniques for probing their domain structures (*19-22*).

Rubidium iron dimolybdate, RbFe(MoO$_4$)$_2$, is a prototypical ferroaxial system with a transition temperature T$_C$ of about 190K(*23-26*). Above T$_C$, the material crystallizes in a high-symmetry para-axial state of space group *P$\bar{3}$m1*. Cooling through the transition temperature spontaneously breaks multiple mirror symmetries, lowering the crystal symmetry to *P$\bar{3}$*. The spontaneous symmetry breaking in RbFe(MoO$_4$)$_2$ results in two opposite ferroaxial domains, A+ and A-, corresponding to clockwise and counterclockwise rotations of the FeO$_6$ octahedra, respectively (Fig. 1a).

This type of symmetry breaking is invisible to linear optical probes, but gives rise to a circular dichroism in electric quadrupole second-harmonic generation (SHG-CD)(*27, 28*), providing a sensitive measure of the ferroaxial domain state(*21*). The emergence of ferroaxial order in RbFe(MoO$_4$)$_2$ is evidenced by temperature-dependent SHG-CD, which displays a sharp increase to a finite value at T$_C$ (Fig. 1b). Notably, the SHG-CD is either positive or negative for the two ferroaxial domains(*21*), enabling direct optical identification of their orientation (see Supplementary Information). Spatial mapping resolves the multi-domain structure below T$_C$, with domain sizes of about 100 micrometers (inset of Fig. 1b)(*20*).

**Ferroaxial conjugate field engineering**

The ferroaxial order parameter is a time-reversal invariant axial vector(*4*), defined as $\vec{A} = \vec{r} \times \vec{P}$, the cross product of the position vector $\vec{r}$ and the local electric dipole moment $\vec{P}$. Direct external control of this order requires a conjugate field that transforms in the same way as the order parameter under all symmetry operations of the system(*29*). This can be realized by the cross product of two different polar fields, $\vec{X} \times \vec{Y}$, applied simultaneously along two different directions(*29-32*). However, combining conventional fields such as electric fields $\vec{E}$, thermal gradients $\nabla T$ or longitudinal strain gradients $\partial_i \varepsilon_{ii}$ in a well-controlled configuration is experimentally challenging(*33, 34*).

We construct a ferroaxial conjugate field $\vec{F} = \vec{Q} \times \vec{E}$ by a terahertz (THz) light field $\vec{E}$ and the displacement $\vec{Q}$ of an infrared-active phonon resonantly driven by this field(*35-48*). Linearly polarized THz pulses, which are commonly used to excite atomic motions parallel to the THz electric field(*41, 43, 49*), do not couple to the ferroaxial order, because the field $\vec{F} = \vec{Q} \times \vec{E}$ vanishes. On the other hand, circularly polarized THz pulses drive circular atomic motions when resonantly exciting a pair of degenerate phonon modes(*50-56*). In this dynamical setting, both the electric field and the phonon displacement rotate continuously, while the cross product of the two fields $\vec{F} = \vec{Q} \times \vec{E}$ remains directionally fixed (Fig. 2a). This configuration offers an experimentally accessible conjugate field to ferroaxial order (*57*), which can be reversed depending on the helicity of the THz electric field.

In RbFe(MoO$_4$)$_2$, the ferroaxial order develops along the crystal *c* axis and is characterized by the axial mode $Q_A$ subject to a temperature-dependent potential $U(Q_A, T)$. Dynamically, the axial mode couples to the z-component of the conjugate field $(\vec{Q} \times \vec{E})_z$ via the interaction potential $\Delta U = \alpha(\vec{Q} \times \vec{E})_z Q_A$. Below T$_C$, the potential $U(Q_A, T)$ takes the form of a symmetric double-well, with each well corresponding to one of the two ferroaxial domains. A circular phonon excitation lifts the degeneracy between the two domains by lowering the energy of one well relative to the other (Fig. 2b). Above T$_C$, RbFe(MoO$_4$)$_2$ is in a para-axial state with a single-well potential. Here, a circular phonon excitation shifts the minimum of the well, transiently inducing as an *axially polarized state* (Fig. 2c). In both cases, the direction of the torque applied on the ferroaxial soft mode is controlled by the helicity of the THz excitation pulse.

**Single-shot switching of ferroaxial order below T_C**

As illustrated in Figure 3a, ferroaxial switching by the THz pulse below $T_C$ is expected due to the transient asymmetric lowering of the double-well potential. This process is captured by the following equations of motion:

$$\frac{\partial^2}{\partial t^2}\vec{Q}_{E_u} + \gamma_{E_u}\frac{\partial}{\partial t}\vec{Q}_{E_u} + \omega_{E_u}^2\vec{Q}_{E_u} = Z_{E_u}^*\vec{E}(t) \qquad (1)$$

$$\frac{\partial^2}{\partial t^2}Q_A + \gamma_A\frac{\partial}{\partial t}Q_A + \frac{\partial}{\partial Q_A}U(Q_A) = \alpha(\vec{Q}_{E_u}(t) \times \vec{E}(t))_z \qquad (2)$$

Here, $\gamma_A$ and $\gamma_{E_u}$ are the damping coefficients, $\omega_{E_u}$ the frequency of the driven phonon modes. $Z_{E_u}^*$ is the effective charge that couples the in-plane phonon displacement vector $\vec{Q}_{E_u}$ to the pulsed THz electric field $\vec{E}(t) = E_0 \begin{pmatrix} \cos(\omega_{E_u}t) \\ \cos(\omega_{E_u}t+\phi) \end{pmatrix} e^{-\frac{t^2}{2\tau^2}}$, where $\phi = \pm\frac{\pi}{2}$ correspond to left-circular and right circular polarization, respectively. Numerical solutions of these equations reveal the evolution of the ferroaxial order as a function of the THz pulse excitation fluence, showing that a single pulse induces a reversal of the order parameter above a switching threshold (Fig. 3b), which is analogous to the coercive field of a ferroelectric. Similar switching behavior is obtained when starting in the opposite domain by changing the helicity of the THz excitation pulse (see Supplementary Information).

We validated this approach using the optical setup illustrated in Figure 4a. We excited the doubly degenerate $E_u$-symmetry phonon at 24 THz frequency in the *ab*-plane of $RbFe(MoO_4)_2$ with resonant, circularly polarized THz pulses, and probed the ferroaxial state with SHG-CD measurements (see Supplementary Information). First, we applied a single right-circularly polarized THz pulse with a fluence of 20 mJ/cm² at normal incidence to an A+ domain at 180 K lattice temperature (below $T_C$), identified by its positive static SHG-CD. Following this single-pulse excitation, the static SHG-CD at the excited sample spot was measured again and showed a negative static SHG-CD of approximately same magnitude, indicating that the A+ state was locally switched to an A- state (Fig. 4b). In a second step, a single left-circularly polarized THz pulse, with the same fluence but opposite helicity, illuminated the same spot. Following this second pulse, the static SHG-CD reverted to a positive value, evidencing that the system switched back to the A+ state.

A sequence of THz pulses with alternating helicities was then applied to the same sample spot, with static SHG-CD measurements performed between pulses to monitor the ferroaxial state. The signal was observed to alternate between positive and negative values, demonstrating reversible optical switching between A+ and A- ferroaxial states (Fig. 4b).

We investigated the fluence dependence of the ferroaxial state switching, aiming to map a 'hysteresis loop' of the ferroaxial order under circular phonon excitation. The experiment was performed with single, right-circularly polarized THz pulses of increasing fluence, with static SHG-CD measured for about 20 minutes after each excitation pulse. As shown in Figure 5a, single THz pulses up to an excitation fluence of 12.5 mJ/cm$^2$ left the A+ ferroaxial state unaffected with a positive static SHG-CD. A subsequent THz pulse with a fluence of 14 mJ/cm$^2$ inverted the static SHG-CD signal to a negative value, evidencing the ferroaxial state switching above this threshold. On the same sample spot, now in the A- ferroaxial state, this protocol was repeated using left-circularly polarized THz pulses, i.e. of opposite helicity. The SHG-CD signal returned to a positive value, characteristic of the A+ state, at the same excitation fluence of 14 mJ/cm$^2$ (Fig. 5b). The observed threshold behavior is in qualitative agreement with the predictions from the equations of motion in our circular phonon excitation model (Fig. 3b).

Notably, the rewritable domain was found to be virtually non-volatile, remaining in the switched state for more than six hours – the duration between switching from A+ state to A- state and back to the A+ state in the measurement shown Fig. 5b.

**Polarizing the para-axial state above $T_C$**

We further explored the possibility of using circular phonon excitation to induce a transient axial polarization in para-axial RbFe(MoO$_4$)$_2$ above $T_C$. A left-circularly polarized THz pulse centered at 24 THz, as above in resonance with the in-plane infrared-active $E_u$-symmetry phonons, is expected to transiently shift the single-well potential of the para-axial state, centered at $Q_A = 0$, to an A+ axial state (Fig. 6a). Conversely, a transient A- axial state is expected for excitation with a right-circularly polarized THz pulse.

Figure 6b shows the time-resolved SHG-CD induced by excitation with a 600-femtosecond, left-circularly polarized THz pulse at a fluence of 6 mJ/cm² at 200 K (above $T_C$). The short-lived, positive SHG-CD signal refers to the emergence of a transient A+ axial polarization. Its lifetime suggests that the duration of the axial conjugate field is determined by the temporal overlap between the THz pulse and the resonantly driven phonon. Excitation with a right-circularly polarized THz pulse induced a short-lived negative SHG-CD signal (Fig. 6c), evidencing a transient A- axial polarization. The helicity-dependent response above $T_C$ is consistent with a model in which circular phonon excitation serves as the conjugate field for axial order. Excitation linearly polarized THz pulses, did not show detectable SHG-CD, confirming that the conjugate field requires circularly polarized excitation (see Supplementary Information).

The temperature dependence of the photo-induced axial state above $T_C$ is shown in Fig. 6d. The amplitude of the transient SHG-CD signal, measured under left-circularly polarized excitation at a fixed fluence of 6 mJ/cm², increases as the sample temperature approaches $T_C$. This response of the axial order parameter is characteristic of ferroic systems, which exhibit increased susceptibility near phase transitions(*58, 59*).

The frequency dependence of the photo-induced axial state as a function of the center frequency of the THz excitation pulses, measured at 200K and a constant excitation fluence of 6 mJ/cm², is shown in Fig. 6e. We found a resonant enhancement in the transient SHG-CD response at the 24-THz frequency of the doubly degenerate in-plane $E_u$-symmetry phonon mode, which verifies the essential role of resonant phonon excitation when inducing the axial order by light (see Supplementary Information).

**Discussion and conclusion**

We experimentally demonstrated the generation of a conjugate field for ferroaxial order(*31*) in the prototypical ferroaxial system RbFe(MoO$_4$)$_2$. This protocol is expected to be broadly applicable across a wide range of ferroaxial materials(*31, 60-62*). Switching the ferroaxial order represents a distinct advantage over earlier attempts to reverse ferroelectric polarization in LiNbO$_3$ with light, where excitation of an auxiliary high-frequency infrared-active mode was nonlinearly coupled to the ferroelectric soft mode(*38, 63*). In that case, a complete reversal of the ferroelectric polarization was inhibited, likely due to the buildup of a depolarizing field at the driven region, which destabilized the switched state(*64*). The control of the ferroaxial state (*65*) opens opportunities for applications in data storage and next-generation electronic and magnetic devices(*62, 66, 67*), flanking the field of optically controlled ferro- and ferrimagnetic order(*68-74*). Moreover, the emergence of axiality in the para-axial state above the critical temperature highlights the strength of coherent electromagnetic fields to engineer functionality of complex materials that do not have equilibrium counterparts.

Figures:

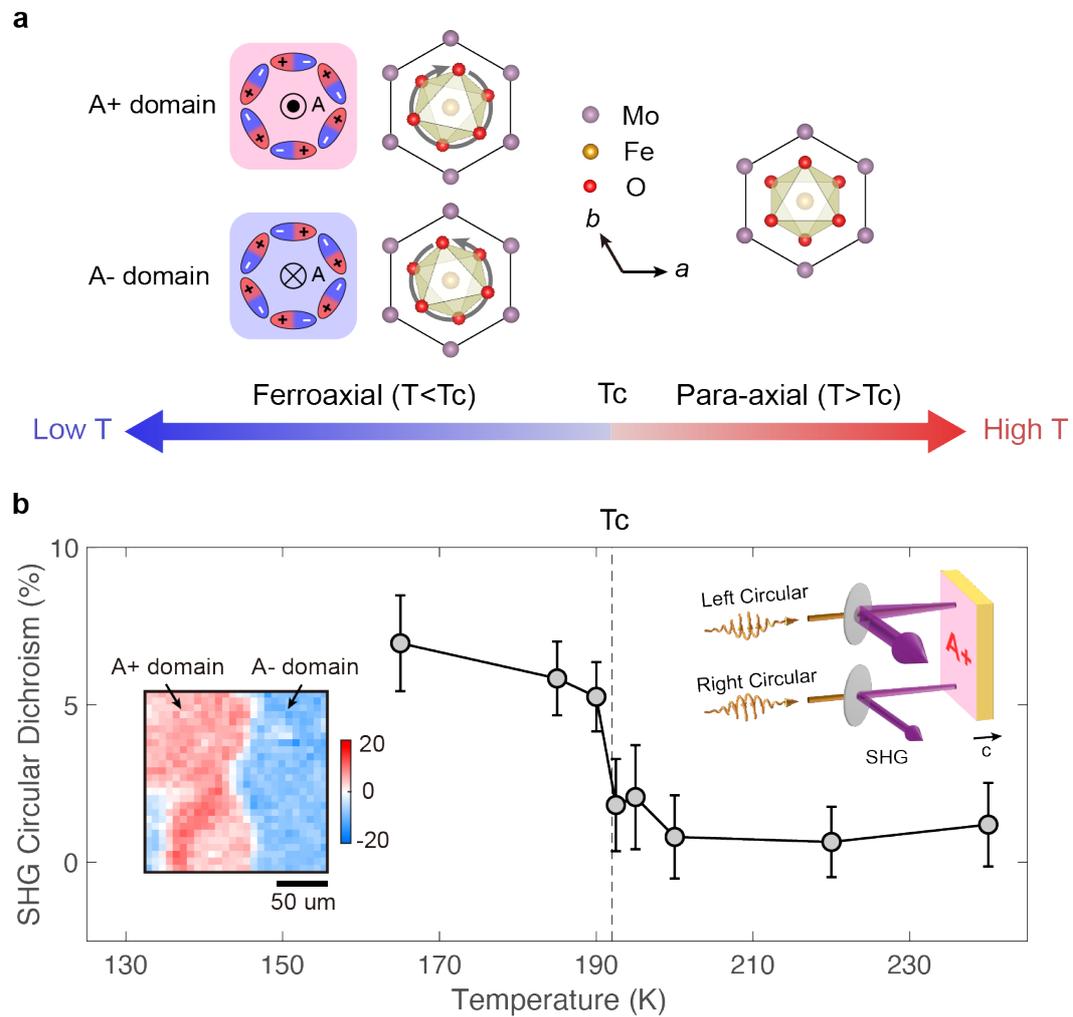

**Fig. 1 | Ferroaxial order in RbFe(MoO₄)₂. a,** Crystal structure of RbFe(MoO₄)₂ viewed along the crystal $c$ axis, above and below the structural phase transition temperature $T_C$. Below $T_C$, two ferroaxial domains exist (A+ and A-), corresponding to clockwise and counterclockwise rotations of the FeO₆ octahedra, respectively. **b,** Ferroaxial transition characterized by second-harmonic generation circular dichroism (SHG-CD). Error bars denote the standard error of the mean. Right inset: schematic of the SHG-CD measurement. In the A+ ferroaxial domain, the SHG intensity is higher for the left-circularly polarized probe than for the right-circularly polarized probe. Left inset: SHG-CD domain mapping at 170K (below Tc).

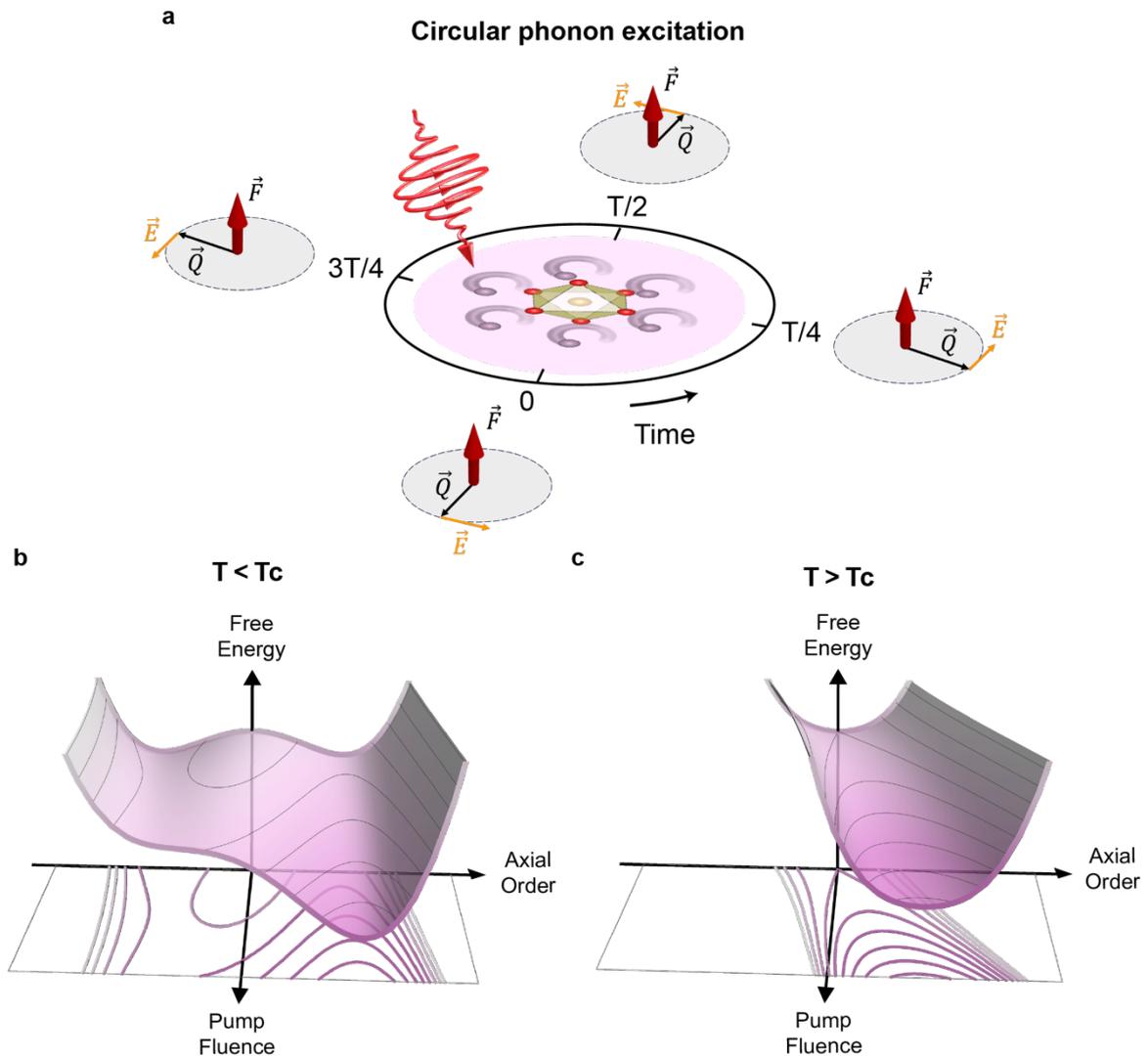

**Fig. 2 | Circular phonon excitation as a conjugate field of axial order. a,** A circularly polarized THz pulse, resonant with a doubly degenerate infrared-active phonon, excites circular phonon motion. At each moment during the drive, the electric field $\vec{E}$ and the phonon displacement $\vec{Q}$ are orthogonal, and the direction of the cross product of $\vec{E} \times \vec{Q}$ remains fixed. **b,** Potential energy landscape of axial order under circular pump below $T_C$. A circularly polarized THz pump biases the double-well potential, favoring one domain over the other, as shown here for a left-circularly polarized THz pulse. **c,** Potential energy landscape of axial order under circular pump in the para-axial state above $T_C$. A circularly polarized pump shifts the potential energy minimum from zero to a finite value of the axial order parameter, shown here again for a left-circularly polarized THz pulse.

## Single-shot switching below Tc

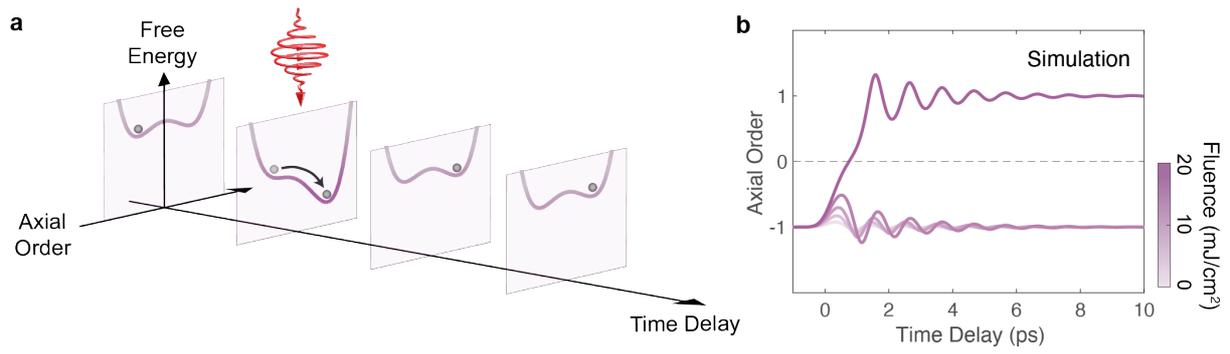

**Fig. 3 | Simulation of single-shot switching of ferroaxial order. a,** Schematic potential energy landscape snapshots of the axial order for circularly polarized phonon excitation below $T_C$. The circularly polarized pump biases the double-well potential and switches the ferroaxial domain. **b,** Dynamics of the axial order for this excitation at various fluences below and above threshold.

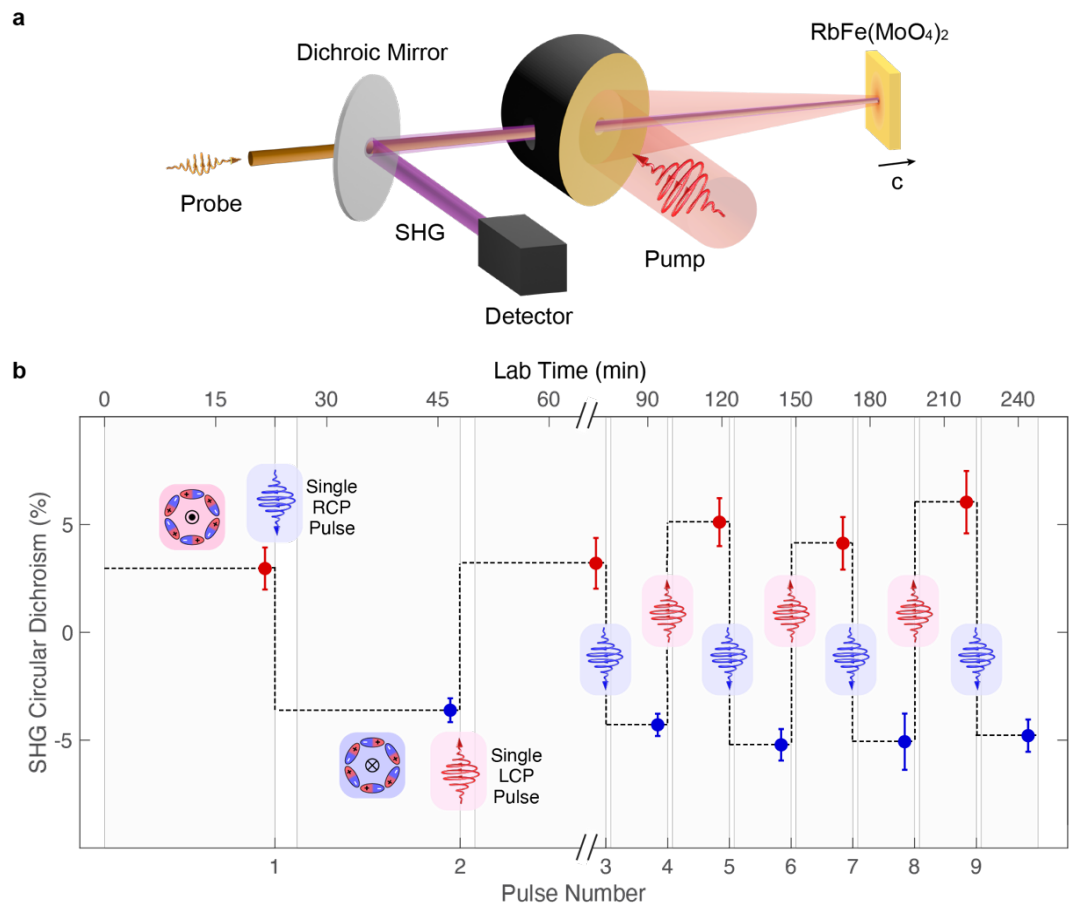

**Fig. 4 | Demonstration of single-shot switching of ferroaxial order below $T_C$. a,** Schematic of the experimental setup. A circularly polarized THz excitation pulse, polarized in the *ab*-plane, drives circular motion of the doubly degenerate $E_u$ symmetry phonons in $RbFe(MoO_4)_2$. SHG-CD monitors the ferroaxial state at the excited position. **b,** Measurement of the ferroaxial domain state between single THz excitation pulses. The SHG-CD measurement averages over the shaded time windows. Error bars denote the standard error of the mean.

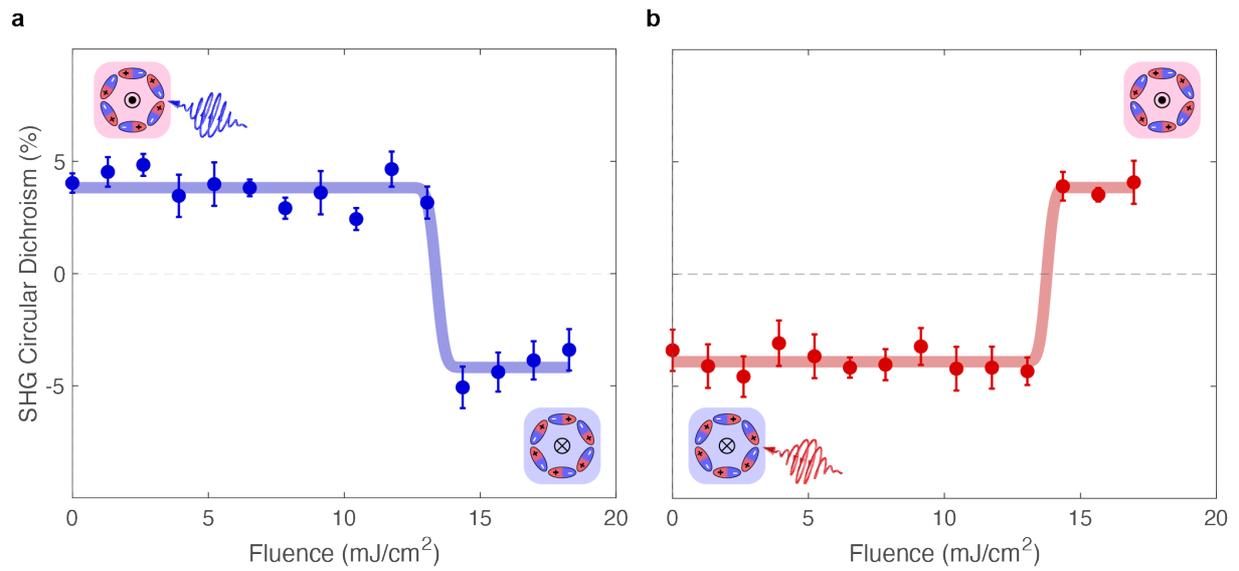

**Fig. 5 | Fluence-dependent single-shot switching of the ferroaxial order. a,** Measurement of the SHG-CD, probing the ferroaxial order after each of the single, right-circularly polarized THz excitation pulses of increasing fluence. **b,** Same measurement for left-circularly polarized single-shot pump pulses. Error bars denote the standard error of the mean.

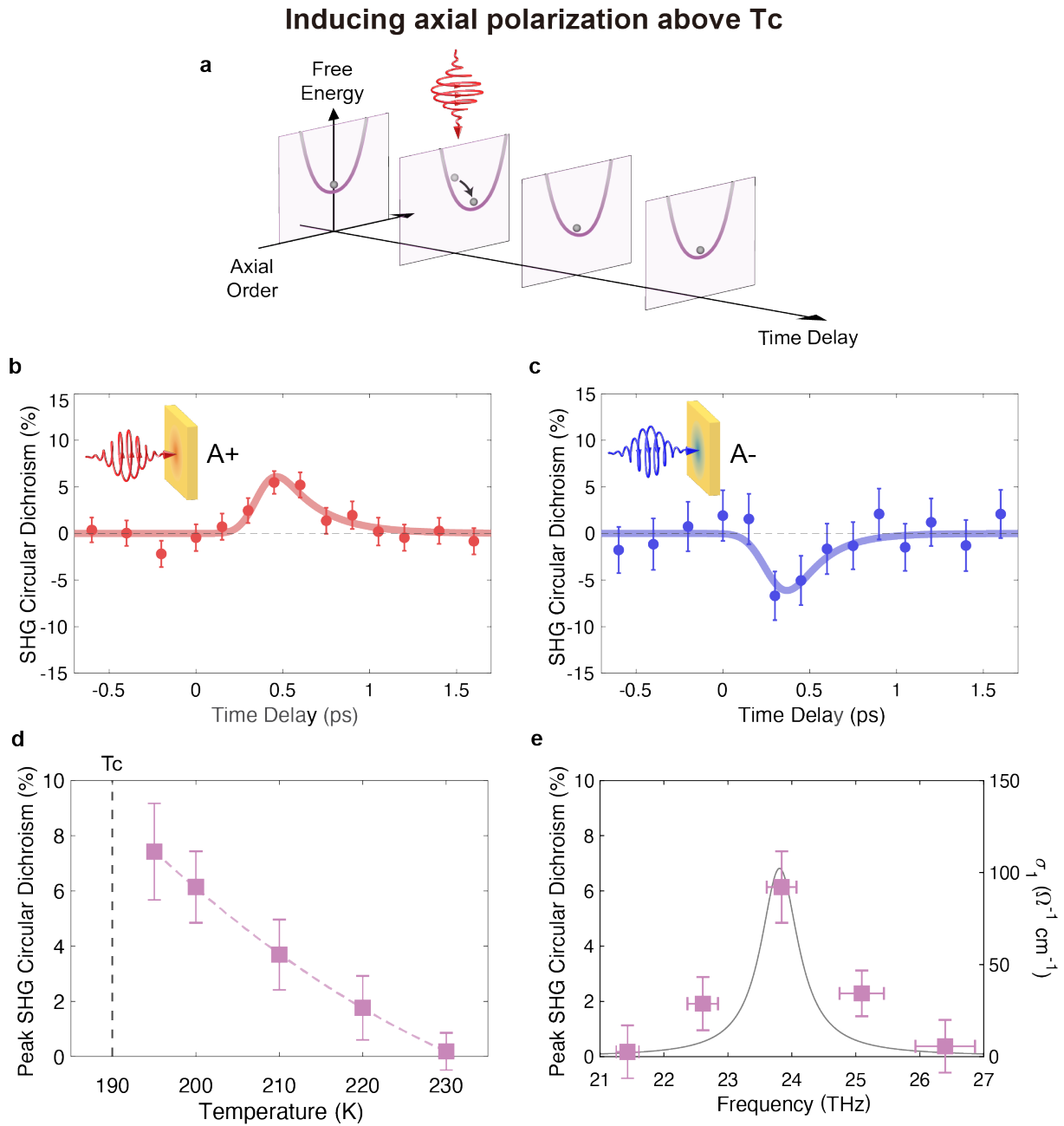

**Fig. 6 | Photo-induced axial order in the para-axial state above T$_C$. a,** Schematic potential energy landscape snapshots of the axial order for a circularly polarized THz excitation pulse above Tc, shifting the potential energy minimum. **b,** Time delay dependent SHG-CD measurement for right-circularly polarized excitation. **c,** Time delay dependent SHG-CD measurement for left-circularly polarized excitation. **d,** Peak amplitude of the transient SHG-CD response as a function of temperature above T$_C$. **e,** Peak amplitude of the transient SHG-CD response as a function of the center frequency of the excitation pulse. The gray curve corresponds to the real part of the optical conductivity. Vertical error bars denote the standard error of the mean. Horizontal error bars in **e** represent the 1$\sigma$ confidence interval of the center frequency of the excitation pulses.

# Supplementary Materials for

# Ultrafast non-volatile rewritable ferroaxial switching

Z. Zeng[1,2], M. Först[1], M. Fechner[1], D. Prabhakaran[2], P. G. Radaelli[2], A. Cavalleri[1,2]

[1]*Max Planck Institute for the Structure and Dynamics of Matter, Hamburg, Germany*

[2]*Department of Physics, Clarendon Laboratory, University of Oxford, Oxford, United Kingdom*

*S1. Materials and Methods*

*S2. SHG Circular Dichroism and the Assignment of Ferroaxial Domains*

*S3. Density Functional Theory Calculations*

*S5. Simulations of the Axial Order Dynamics*

*S5. Time-resolved SHG-CD above $T_c$ for Linearly Polarized Excitation*

## S1. Materials and Methods

### S1.1. Experimental Setup

A schematic drawing of the pump-probe setup used in the experiment is shown in Figure S1. The THz pump pulses were generated by difference frequency generation (DFG) in a GaSe crystal(*75*), using the independently wavelength-tunable near-infrared signal outputs of two optical parametric amplifiers (OPAs). The OPAs were seeded by the same white light and pumped by overall 11-mJ, 35-fs pulses at 800 nm wavelength from a Ti:sapphire amplifier operating at 1-kHz repetition rate. Following the DFG process, linearly polarized THz pump pulses were temporally stretched to ~600 fs by passing through two 5-mm-thick ZnSe crystals positioned at the Brewster angle. The THz spectrum was characterized by Fourier transform infrared spectroscopy (Fig. S2a).

A CdSe-based infrared quarter waveplate induced circular polarization of the THz pulses, which was characterized by detecting the optical power transmitted through a rotating linear polarizer (Fig. S2b). A combination of two free-standing wire grid polarizers before the quarter waveplate allowed us to reverse the helicity of the THz excitation pulses between left- and right-circularity and control the excitation fluence without rotating the quarter waveplate. This approach ensured the identical alignment and spot position on the sample for both helicities. An off-axis parabolic mirror was then used to focus the THz beam onto the sample with a ~70 µm spot size (full width at half maximum intensity, FWHM).

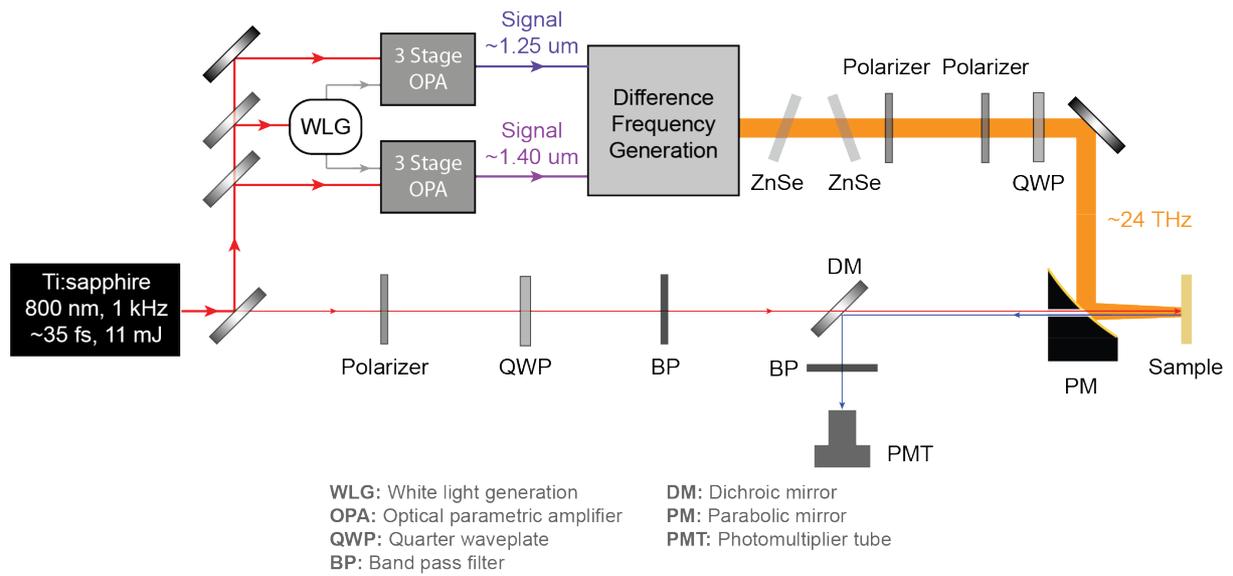

**Figure S2. | Schematic of the experimental setup.**

The probe pulses for the measurement of the second-harmonic generation circular dichroism (SHG-CD) were sourced by the same amplifier system and focused with a lens onto the sample to a spot size of ~30 μm (FWHM). Circular polarization of the probe incident on the sample was achieved using another quarter waveplate. The SHG-CD signals were measured in reflection geometry with a coherence length in the SHG process of about 50 nm,(*76*) shorter than the 500-nm extinction depth of the THz pulses at the phonon resonance. A pair of narrow bandpass filters was employed before and after the sample to filter out side-band components in the pump-probe measurements arising from electric-field-induced second-harmonic generation (EFISH) effect. The filter before the sample transmits at 808 nm center wavelength, with a 5-nm bandwidth, whilst the filter on the SHG output is centered at 404 nm with a bandwidth of 2 nm.

The pump and probe beams were aligned collinearly and at normal incidence to the sample surface of c-cut RbFe(MoO$_4$)$_2$, in this way preserving the circular polarization in the excitation and detection processes. All measurements were performed in an optical cryostat under vacuum conditions (< 10$^{-6}$ mbar).

For the preparation of single THz excitation pulses, two optical choppers with 10% duty cycle, synchronized to the Ti:sapphire amplifier system, were placed in the THz beam path before the sample, reducing the repetition rate from 1 kHz to 10 Hz. A synchronized fast shutter was then used to select single pulses for the optical excitation.

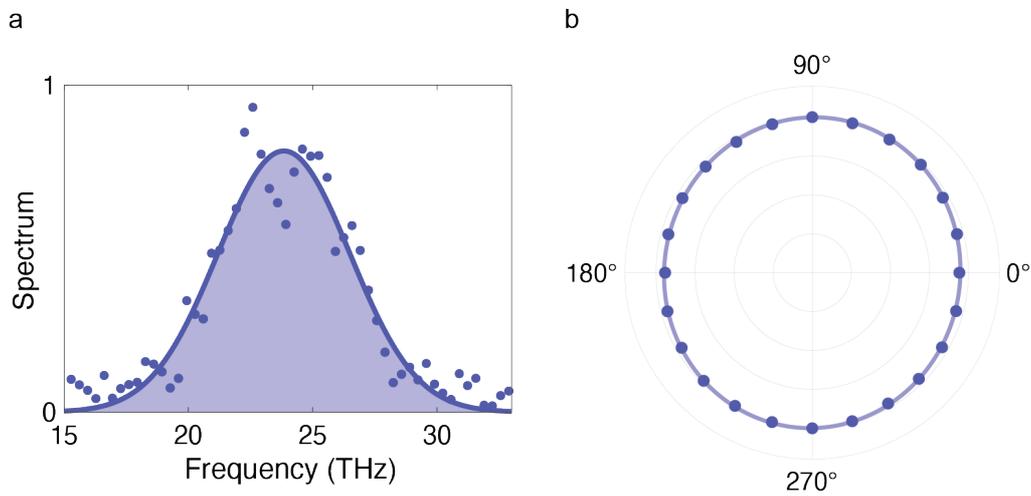

**Figure S2. | Characterization of the THz excitation pulses.** (a) Frequency spectrum of the excitation pulse, centered at 24 THz, characterized by Fourier transform infrared spectroscopy. (b) Measurement of the polarization state of the left-circularly polarized 24-THz excitation pulses, showing a 97% ellipticity.

## S1.2. Sample Preparation and Characterization

Single crystal RbFe(MoO$_4$)$_2$ was grown using the flux method and characterized by x-ray diffractometry. A sample with an optically flat c-cut surface and thickness of ~200 μm was used in this experiment. In the experiments, it was cooled and heated with a slow rate of 0.1 K/min near the 190-K critical temperature.

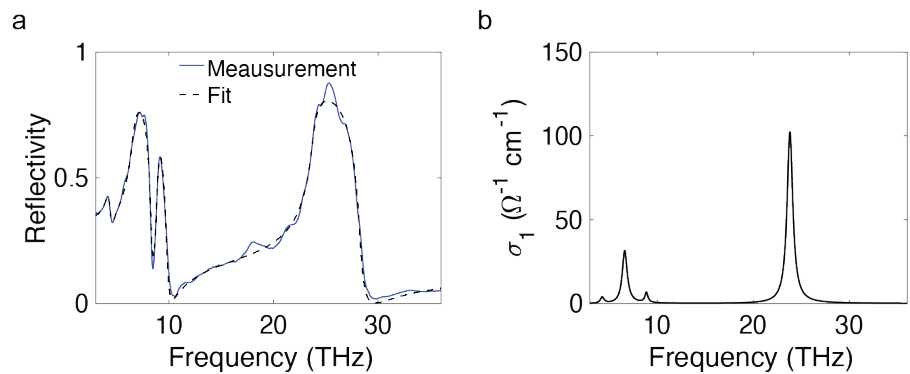

Figure S3. | Static optical characterization of the RbFe(MoO$_4$)$_2$ sample. (a) Measured reflectivity spectrum in the THz range with the electric field polarized in the crystal *ab*-plane, along with the corresponding fit. (b) Frequency dependent real part of the optical conductivity extracted from the fit.

The static optical properties in the THz frequency range were characterized by Fourier-transform infrared reflectivity (FTIR) measurements, with the light electric field polarized in the crystal *ab*-plane. As shown in Figure S3, four E$_u$-symmetry optical phonon modes of RbFe(MoO$_4$)$_2$ were found in the reflectivity spectrum, which was fitted using the following dielectric function for infrared-active phonons(*77*)

$$\varepsilon(\omega) = \varepsilon_\infty + \varepsilon_\infty \sum_j \frac{\omega_{LO,j}^2 - \omega_{TO,j}^2}{\omega_{TO,j}^2 - \omega^2 - i\omega\gamma_j}$$

The fitting results are summarized in Table S1. In the experiments reported here, we resonantly excited the E$_u$(4) mode at 24 THz.

| Phonon Mode | $\omega_{TO,j}$ (THz) | $\omega_{LO,j}$ (THz) | $\gamma_j$ (THz) | $\varepsilon_\infty$ |
|---|---|---|---|---|
| E$_u$(1) | 4.3 | 4.8 | 0.45 | / |
| E$_u$(2) | 6.6 | 9.9 | 0.61 | / |
| E$_u$(3) | 8.9 | 9.2 | 0.38 | / |
| E$_u$(4) | 23.8 | 27.8 | 0.72 | / |
| / | / | / | / | 4.01 |

Table S1. | Results of fitting the infrared reflectivity spectrum, polarized in the a*b* plane, by the E$_u$-symmetry phonon modes in RbFe(MoO$_4$)$_2$.

## S2. SHG Circular Dichroism and the Assignment of Ferroaxial Domains

RbFe(MoO$_4$)$_2$ crystallizes in a centrosymmetric structure both above and below T$_C$. Since electric dipole SHG signal is forbidden, the observed SHG signal arises from higher-order processes, predominantly electric quadrupole SHG(*25*).

The SHG Circular Dichroism (SHG-CD) signal in this study is defined as the normalized difference of the SHG intensity detected between left-circularly polarized (LCP) and right-circularly polarized (RCP) normal-incident near-infrared probe pulses.

$$SHG\text{-}CD = \frac{I_{LCP} - I_{RCP}}{I_{LCP} + I_{RCP}}$$

Based on symmetry analysis, the SHG-CD signal is expected to be zero in the para-axial state and finite in the ferroaxial state, with opposite signs for opposite ferroaxial domains(*21*).

The exact correspondence between the sign of the SHG-CD signal and the ferroaxial domain state is unknown and not easily accessible through static characterization. For the convenience of analysis and discussion, we assigned domains with a positive SHG-CD signal as A+ domains and those with a negative SHG-CD signal as A- domains. The validity of this assignment is supported by the helicity-dependent response induced by the circular THz excitation, which agrees in sign with theoretical calculations(*57*).

**S3. Density Functional Theory Calculations**

We carried out first-principles calculations based on density functional theory (DFT+U) to investigate the phonon excitation spectrum, anharmonic lattice interactions, and optical properties of RbFe(MoO$_4$)$_2$. All simulations were conducted using the Vienna Ab-initio Simulation Package (VASP, version 6.5.1)(*78-80*), with phonon-related calculations carried out via the Phonopy package(*81*).

Pseudopotentials were used within the Projector Augmented Wave (PAW) framework(*82*), employing standard potentials for Rb (4p$^6$5s$^1$), Fe (3p$^6$3d$^6$4s$^2$), Mo (4d$^5$5s$^1$), and O (2s$^2$2p$^4$). The Local Density Approximation (LDA) was chosen for the exchange-correlation functional, augmented by the Hubbard *U-J* correction to account for the localized d-electrons of Fe.

Consistent with a previous study(57), we found that the phonon and related properties are insensitive to the precise value of $U$, and thus adopted $U = 4$ eV and $J = 0$ eV.

After convergence testing, structural relaxation and phonon calculations were performed using a 7×7×5 Monkhorst-Pack k-point grid(83) and a plane-wave cutoff energy of 600 eV. Self-consistent iterations continued until total energy changes fell below $10^{-8}$ eV. All phonon and anharmonic coupling constant evaluations were carried out in the primitive unit-cell of $RbFe(MoO_4)_2$.

Our starting point was the $P\bar{3}m1$ high-temperature crystal structure of $RbFe(MoO_4)_2$, which we fully structurally relaxed. The resulting lattice parameters are listed in Table S2. Phonon calculations for this structure revealed an unstable $A_{2g}$ ferroaxial soft mode with an imaginary frequency. The frequency of the driven doubly degenerate $E_u$ phonon mode is also reported in Table S2.

To explore the ferroaxial low-temperature phase, we applied a frozen-phonon approach by displacing the structure along the unstable $A_{2g}$ mode. This resulted in the double-well potential shown in Figure S4, indicating a structural instability toward a lower-symmetry $P\bar{3}$ ground state.

| Physical Properties | Parameter/Site | a (Å) / x | b (Å) / y | c (Å) / z | Frequency (THz) |
|---|---|---|---|---|---|
| Lattice constants | / | 5.65 | 5.65 | 7.20 | / |
| Structural coordinates | Rb | 0 | 0 | 1/2 | / |
| | Fe | 0 | 0 | 0 | / |
| | 2Mo | 1/3 | 2/3 | 0.240 | / |
| | 2Oa | 1/3 | 2/3 | 0.482 | / |
| | 6Op | 0.160 | 0.840 | 0.162 | / |
| **Phonon** | $E_u(4)$ | / | / | / | 24.42 |

Table S2. | Relaxed lattice parameters and selected phonon mode frequency of high-symmetry $RbFe(MoO_4)_2$.

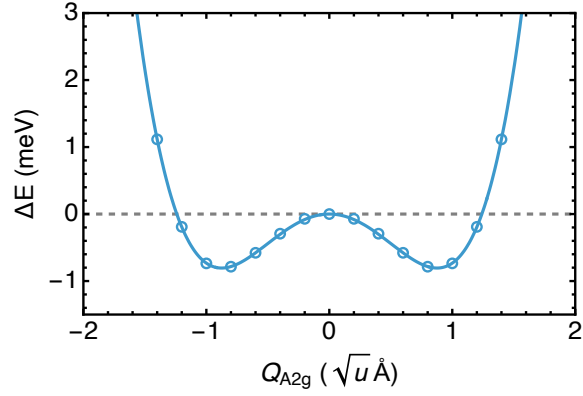

**Fig S4. | Double-well potential energy profile obtained by displacing the high-symmetry RbFe(MoO₄)₂ structure along the unstable A$_{2g}$ ferroaxial mode. The two minima correspond to symmetry-broken configurations associated with the low-temperature $P\bar{3}$ phase, indicating a structural phase transition driven by the soft mode.**

From the relaxed high-temperature structure, we mapped the coupling between the infrared-active mode at 24 THz and the optically silent A$_{2g}$ soft mode. Specifically, we computed the off-diagonal term of the mode effective charge tensor as a function of the A$_{2g}$ displacement, which characterizes the trilinear coupling term $\alpha(\vec{Q} \times \vec{E})_z Q_A$ following the methodology outlined in a previous study(*57*). The resulting coupling coefficient in our computations is $\alpha = 0.026$ q$_e$/uÅ, which results in a tilted energy landscape for axial order under circularly polarized excitation (Fig. S5).

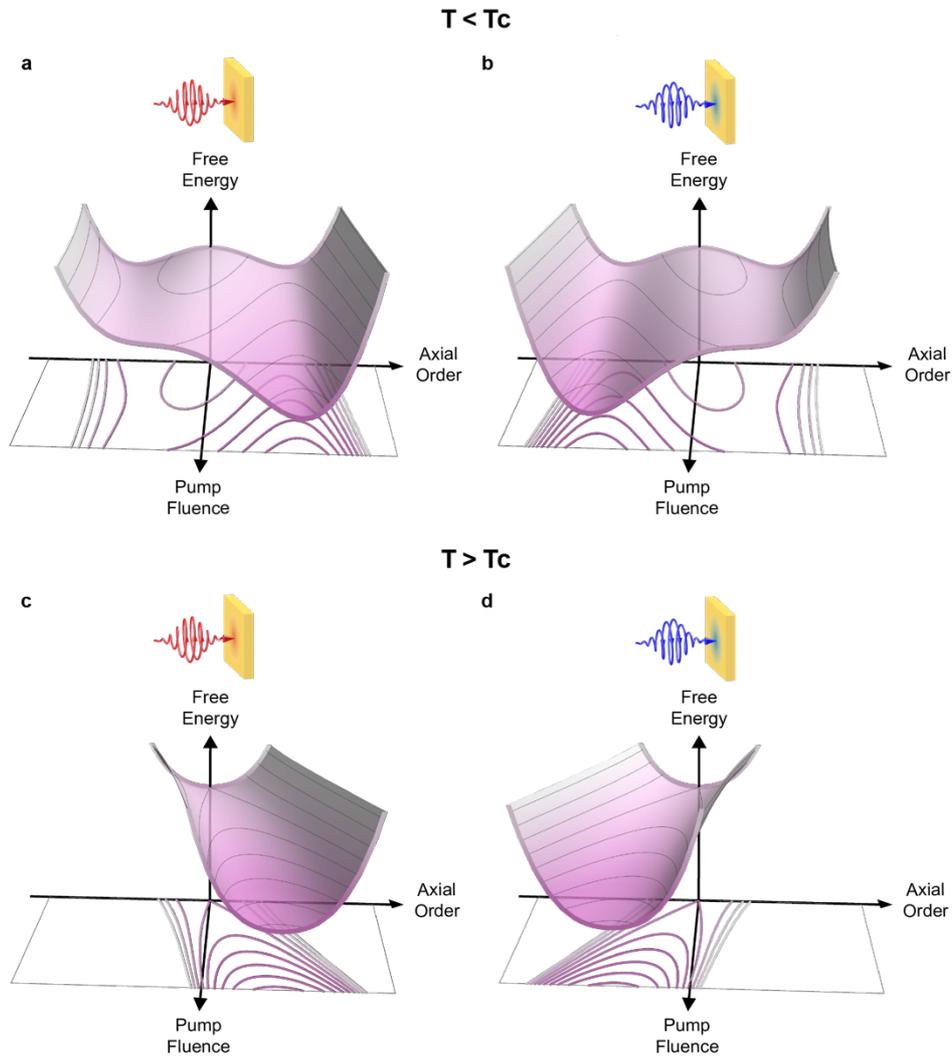

**Figure S5. | Schematic potential energy landscape of axial order for circularly polarized THz excitation.** (a) Axial order below $T_C$ with left-circularly polarized THz excitation. (b) Axial order below $T_C$ with right-circularly polarized THz excitation. (c) Axial order above $T_C$ with left-circularly polarized THz excitation. (d) Axial order above $T_C$ with right-circularly polarized THz excitation.

## S4. Simulations of the Axial Order Dynamics

To model the dynamics of the driven axial state and ferroaxial switching, we solved the equations of motion for the infrared-active mode coupled to the axial soft mode (see main text). We used a circularly-polarized Gaussian pulse that replicates the experimental conditions with a pulse duration of 600 fs (FWHM). To account for finite temperature effects, the harmonic term of the $A_{2g}$ soft mode was scaled using experimentally determined mode frequencies in the low-temperature phase (Fig. S6)(*84*), following the fitting function

$$\omega^2(T) = \omega_0^2 \left|1 - \left(\frac{T}{T_C}\right)^\beta\right|$$

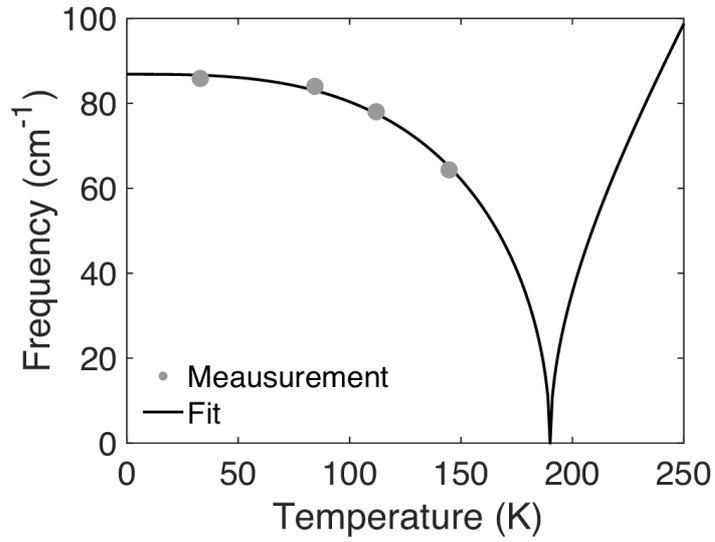

**Figure S6. |** Experimentally determined axial soft mode frequency (grey dots) and a fit to the data.

The simulated dynamics as a function of fluence below $T_C$ are shown in Figure S7. For the A- ferroaxial initial state, a left-circularly polarized THz excitation pulse above a certain fluence threshold switches the state to the opposite A+ state (Fig. S7a), while the right-circularly polarized excitation leaves it in the A- ferroaxial state (Fig. S7b). Conversely, for the A+ ferroaxial initial state, switching to the A- state is achieved with a right-circularly polarized excitation pulse above the threshold (Fig. S7c), while the left-circularly polarized excitation leaves the A+ state unchanged (Fig. S7d).

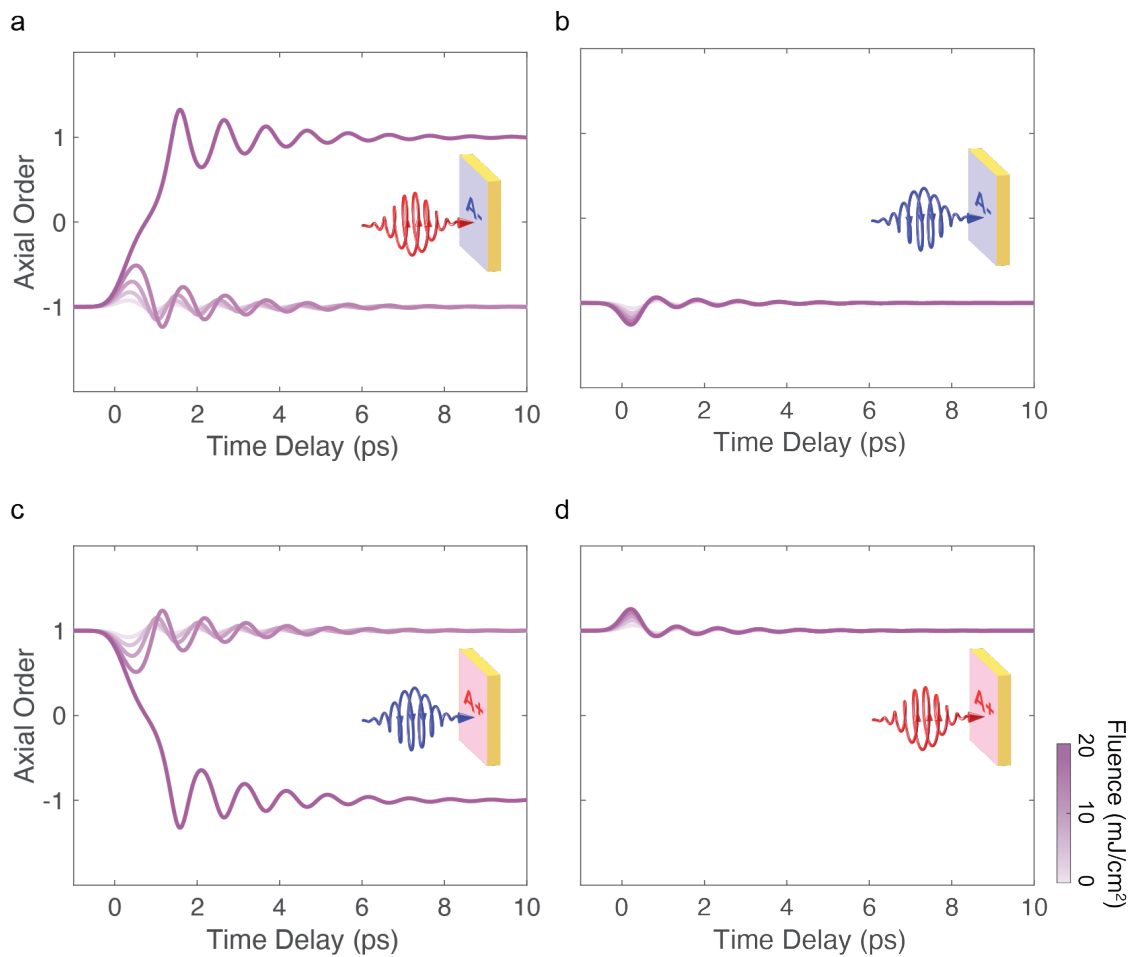

**Figure S7. | Calculated dynamics of axial order for circularly polarized excitation at different fluences below $T_C$. (a) Dynamics of an A- ferroaxial initial state under left-circularly polarized THz excitation. (b) Dynamics of an A- ferroaxial initial state under right-circularly polarized THz excitation. (c) Dynamics of an A+ ferroaxial initial state under right-circularly polarized THz excitation. (d) Dynamics of an A+ ferroaxial initial state under left-circularly polarized THz excitation.**

Simulations were also performed above $T_C$, for a fixed excitation fluence of 6 mJ/cm² matching the experimental conditions. A left-circularly polarized pump induces a transient positive axial order (Fig. S8a), while a right-circularly polarized pump induces a transient negative axial order (Fig. S8b). In both cases, the amplitude of the transient order increases as the temperature approaches the ferroaxial transition temperature, reflecting the softening of the axial mode.

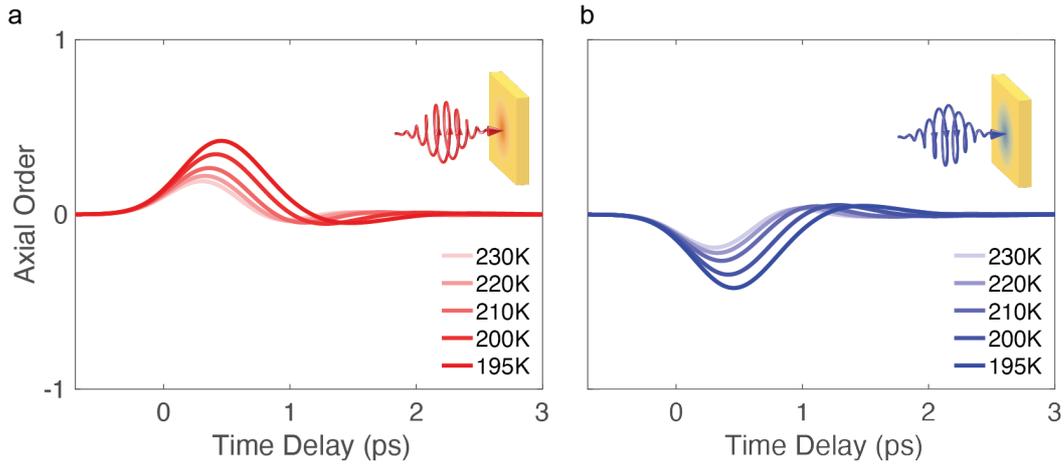

**Figure S8. |** Calculated dynamics of the axial order for circularly polarized excitation pulses at different temperatures above $T_C$. (a) Transient A+ axial order induced by a left-circularly polarized THz pulse. (b) Transient A- axial order induced by a right-circularly polarized THz pulse.

## S5. Time-resolved SHG-CD above $T_C$ for Linearly Polarized Excitation

Time-resolved SHG-CD measurements of a possible light-induced axial state above $T_C$ were also taken using linearly polarized THz pulses at a fluence of 6 mJ/cm² and at 200 K lattice temperature, again with the pump polarization in the crystal *ab*-plane (perpendicular to the crystal *a* axis). We did not observe finite transient SHG-CD, strongly suggesting that the conjugate field of the axial order vanishes under linearly polarized excitation, in agreement with the model presented in the main text.

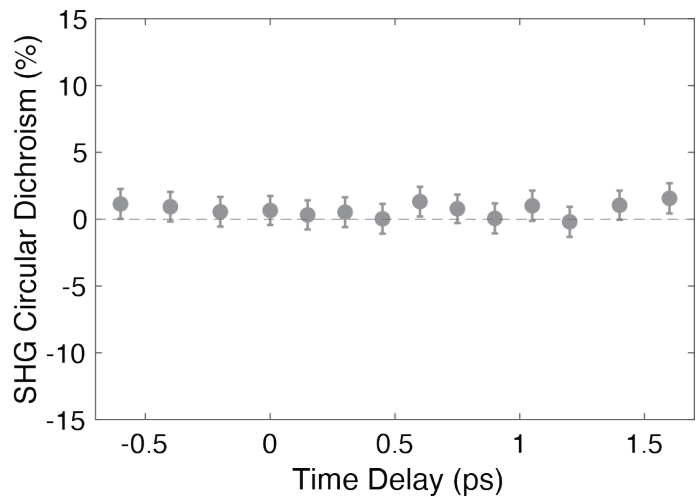

**Figure S9. |** Time-resolved SHG-CD using linearly polarized THz excitation pulses of 6mJ/cm² fluence at 200 K lattice temperature (above $T_C$). Error bars denote the standard error of the mean.